# DYNAMICAL DERIVATION OF BODE'S LAW

R. W. Bass[1] & A. Del Popolo[2]

[1] Prof. of Physics and Astronomy, BYU (1971-1981)

[2] Bogazici University; Physics Department; 34342 Bebek / Istanbul; TURKIYE

**Abstract**

In a planetary or satellite system, idealized as n small bodies in initially coplanar, concentric orbits around a large central body, obeying Newtonian point-particle mechanics, resonant perturbations will cause dynamical evolution of the orbital radii except under highly specific mutual relationships, here derived analytically. In particular, the most stable situation is achieved (in this idealized model) only when each planetary orbit is roughly twice as far from the Sun as the preceding one, as observed empirically already by Titius (1766) and Bode (1778) . Simplifying the problem by reformulating it as a hierarchical sequence of [unrestricted] 3-body problems, in which gravitational interactions are ignored except between the central body and the body of interest and the next outwardly orbitally adjacent body, it is proved that the resonant perturbations from the outer body will destabilize the inner body [& conversely] unless its mean orbital radius is a unique and specific multiple $\beta$ of that of the inner body. In this way a sequence of concentric orbits can each stabilize the next adiacent inner orbit, until only the outermost orbit remains; but it is already tied to the collection of inner orbits by conservation of total angular momentum and so the entire configuration becomes stabilized. Let $\mu = M_{\text{Max}}/M$ denote the ratio of the mass $M_{\text{MAX}}$ of the largest small body to that, M, of the large central body; in our Solar System, $\mu$ is less than $10^{-3}$. Expanding $\beta$ in a power series in $\mu$, the lowest-order terms for distal multiplier $\beta$ and phase shift $\phi$ are found to start with the following universal constants [for $m = 1, 2, 3$]

$$\beta \equiv \beta + \ldots\ldots = 1/\sqrt{(3/2)^{(2/3)} - 1} = 1.794980 + \ldots\ldots, \phi = m(\pi/2) + \ldots$$

which agrees with the observed distal ratios between Jupiter and the Asteroids, and between Saturn & Jupiter, and Uranus & Saturn. In view of Kepler's Law this implies that the stabilizing resonances must be of the type 2:1 or 5:3 or 3:1 (to the extent that only 3-body interactions and only integers not exceeding 5 are considered). Only analytic complexity has inhibited the present theory from being extended to 4-body and 5-body interactions, as a result of which the observed resonant structure of the Solar System [Ovenden & Roy; Molchanov] can be understood dynamically as the inevitable result of resonant perturbational evolution.

**Introduction**

The distribution of planets in the Solar system is not random and their mean distance, from the Sun, when numbered from the centre, form a rough geometric progression:

$$\frac{r_{n+1}}{r_n} \approx 1.75 \tag{1}$$

or

$$r_n = 0.4 + 0.15 \times 2^n, n = -\infty, 1...8 \tag{2}$$

This last relation is known as Titius-Bode law, and it roughly describes the planetary semi-major axes in astronomical units, with Mercury assigned $i = -\infty$, Venus $i = 1$, Earth $i = 2$, etc. Usually the asteroid belt is counted as $i = 4$. The law fits the planets Venus through Uranus quite well, and successfully predicted the existence and locations of Uranus and the asteroids. However, (i) the law breaks down badly for Neptune and Pluto; (ii) there is no reason why Mercury should have $i = -\infty$, rather than $i = 0$, except that it fits better that way; (iii) the total mass of the asteroid belt is far smaller than the mass of any planet, so it is not clear that it should be counted as one. When this relationship was discovered by Titius of Wittenberg in 1766 and published by Bode six years later, it gave good agreement with the actual mean distances of the planets that were then known-Mercury (0.39 AU), Venus (0.72 AU), Earth (1.0 AU), Mars (1.52 AU), Jupiter (5.2 AU), and Saturn (9.55 AU). Uranus, discovered in 1781, has mean orbital distance 19.2 AU, which also agrees. The asteroid Ceres, discovered 1801, has mean orbital distance 2.77 AU, which fills the apparent gap between Mars and Jupiter. However, Neptune, discovered 1846, has mean orbital distance 30.1 AU, and Pluto, discovered 1930, has mean orbital distance 39.5 AU; these are large discrepancies from the positions 38.8 AU and 77.2 AU, respectively, predicted by Bode's law. It is thought that the outer planets do not fit as well due to millions of tears of comet impacts and a looser gravitational hold. Also, Pluto is thought to have been created outside our solar system and the law would therefore not apply to it.

For the next 200 years there has been a debate in astronomy as to whether Bode's Law is simply chance or if it underlines a yet unseen force in the creation of the solar system. A fairly comprehensive history of the law and attempts to explain it up to the year 1971 can be found in Nieto [43]. Some theories of the origin of the solar system have tried to explain the apparent regularity in the mean orbital distances of the planets, arguing that it could not arise by chance, but must be a manifestation of the laws of physics. Some astronomers hold that the deviation of Neptune and Pluto from their predicted positions signifies that they are no longer at their original positions in the solar system. However, since Bode's law is not a law in the usual scientific sense, i.e., it is not universal and invariant, it alone should not be taken as evidence for such a conclusion. Under the assumption that planets need a certain amount of space to form without competing for material, random numbers subjected to computer simulations of early solar systems have been done which show planet spacing that follow similar laws [44]. Despite this, many scientists over the centuries have come up with explanations for the law. One theory, presented by Stuart Weidenchilling and Donald Davis, claims that the planets ended up where they are due to the frictional drag, bringing the smaller planets closer to the sun and "gravitational perturbations from embryonic planets" Under the right conditions there would be favourable locations for planets to form [45]. Another recent theory is that a simple numerical sequence that "combines scale and rotational invariance, the points at which a physical parameter, such as density, reaches a maximum or a minimum will always follow a simple relationship". Using this insight, Francois Graner and Berengere Dubrulle have improved Bode's Law to give better results. They claim that to believe in Bode's Law is to believe in scale invariance in the early solar system [46]. Summarizing most modern arguments concerning the validity of Bode's law can be assigned to one of three broad classes:

1. Attempts to elucidate the physical processes leading to the Bode law. These are based on a variety of mechanisms, including dynamical instabilities in the protoplanetary disk ([47], [48], [49]), gravitational interactions between planetesimals [44], or long-term instabilities of the planetary orbits ([50], [51], [52]).
2. Discussions that ignore physics but try to assess whether the success of the Bode law is statistically significant ([53], [54], [55]). Conclusions go from the Bode law being real rather than artifactual to the contrary. As observed by Hayes & Tremaine [56], these last conclusions are flawed because of some assumptions in the analyses.
3. Discussions of other laws that may influence the spacing of the planets. Many of these involve resonances between the mean motions of the planets, such as [57] (but see [58]), [59], [60].

For what concerns point 3, the commensurability and resonance phenomena of the solar system motion structure - (including planets, asteroids, planet's satellites) - have been object of detailed discussions and experimental examination in the last years. Systematic observations and measurements have been carried out using all available means. A considerable bulk of empiric data for the solar system has been obtained from the "Voyager 2" mission. The understanding of the inevitable resonance character of evolving mature oscillation systems leads to series of interesting concepts about the resonance character of the Solar system motion dynamics. One of the most outstanding among them is Molchanov's hypothesis about the complete resonance character of the large planets in-orbit motion. A.M. Molchanov has noticed that the mean motion of the nine large planets are related approximately by nine linear homogeneous equations ,

$$k_1^{(j)} n_1 + ..... + k_9^{(j)} n_9 = 0, j = 1, ....9 \qquad (3)$$

with integer coefficients:

$$k_1^{(j)}, ......, k_9^{(j)} \qquad (4)$$

The mean motions of the Jovian, Saturn and Uranian satellites are related by similar equations. If asteroids are considered as planets and excluding Pluto, it is established [61] that the planetary distances obey the following regularity:

$$\frac{a_{k+1}}{a_k} \cong 1.75 \pm 0.20 \qquad (5)$$

i.e. the relation of the semi-major orbital axes of neighboring planets is almost constant.

In the present paper, we perform a dynamical derivation of Bode's law. The basic strategy in the following is as follows: the famous "Poincare map" which is now often implemented by physicists using numerical integration [34] will be implemented analytically. In the present context, this means that an arbitrary line-segment will be defined transverse to a periodic generating orbit, and nearby orbits will be followed until this 1-dimensional "surface of section" is intersected a second time. This defines a continuous mapping of the section into itself, of which map the initial periodic solution is now a fixed point. Behavior of nearby orbits can now be studied by considering the simpler problem of iterating this map in a small neighborhood of the fixed point. It will be proved below that no fixed point exists (i.e. no periodic solution exists) unless at $\mu = 0$ the distal ratio defined in the Abstract is precisely $\beta_o$.

## Problem Formulation

Let $x^i \in \mathbb{E}^2$ ($i = 1, 2, 3, \cdots, n$) denote 2-vectors or elements of 2-dimensional real Euclidean space $\mathbb{E}^2$ which represent the positions of $n$ point-particles of masses $M_i$ in orbit around a central body of mass $M_o$ and position $x^o$. Let $v^i \in \mathbb{E}^2$ denote their velocities $\dot{x}^i \equiv dx^i/dt$ with respect to time $t$, where $\dot{} := d/dt$. Let $x \cdot y \equiv (x,y)$ denote the *scalar product* between any two vectors $x$, $y$, and let $\|x\| := (x,x)^{1/2}$ denote the *norm* (length) of any vector.

It is assumed that $\mu_i := M_i/M_o \ll 1$. Now rescale the masses so that $M_o = 1$ and the smaller masses are $\mu_i \ll 1$. Let $\mu_{MAX}$ denote the largest of the small masses, and rewrite them as

$$\mu_i = \mu \cdot \mu_i^o = \mu \cdot \varepsilon_i^o \cdot \mu_{MAX}, \quad 0 \le \mu \le 1, \quad (0 < \varepsilon_i^o \le 1), \tag{6}$$

so that $\mu$ is the *perturbation parameter*. All that will be PROVED here is for $\mu$ 'sufficiently small', though there are good reasons for believing that an analytical continuation (in the manner of Poincaré) can be made all the way from $\mu = 0$ to $\mu = 1$; this is because the **Leray-Schauder Index** of the 'generating solution' at $\mu = 0$ will be *proved* to be unity, and the fact that this *integer-valued* topological invariant, specifying in some sense the true multiplicity of actually existing solutions, is a *continuous* homotopy invariant and so can change its value *discontinuously*, i.e. terminate', ONLY at a bifurcation or singularity of the solution (i.e. a collision or an ejection), demonstrates that the homotopy on $\mu \in [0,1]$ is legitimate *unless* there is an intermediate value of $\mu < 1$ at which there is a collision between two bodies or one body is ejected to infinity; this makes rigorous Strömgren's *Principle of Natural Termination*, discovered empirically by numerical integration [1], in which a family of periodic solution*s* being studied by the variation of a parameter can cease to exist *only* at an ejection/collision event.

Now introduce *relative* coordinates, in which each $x^i$ is replaced by $(x^i - x^o)$, although for convenience the latter will be renamed as $x^i$; it is well known that then the system becomes, for the indices $i = 1, 2, \cdots, n$,

$$\dot{x}^i = v^i, \tag{7a}$$

$$\dot{v}^i = -G \cdot (1 + \mu_i) \cdot (x^i/\|x^i\|^3) - G \cdot \sum_{\substack{j=1 \\ j \ne i}}^{n} \mu_i \cdot \left\{ [x^j/\|x^j\|^3] - [(x^j - x^i)/\|(x^j - x^i)\|^3] \right\}, \tag{7b}$$

where the absolute acceleration of the coordinate system's origin, which now coincides with the position of the Sun or central body, is the source of the first term in the summation (cf. Kurth [24], pp. 84-85), and where $G$ denotes the Newton-Cavendish parameter or gravitational constant.

Now, following Kurth, assume that each planet is affected significantly only by the Sun and the next further outward planet:

$$\dot{v}^k = -G \cdot (1 + \mu_k) \cdot (x^k / \|x^k\|^3) + G \cdot \mu_k \cdot \left[ [(x^m - x^k)/\|x^m - x^k\|^3] - [x^m / \|x^m\|^3] \right],$$

$$k = (i,j), \quad j = (i+1); \quad m = m(k); \quad m(i) = j; \quad m(j) = i; \quad (i = 1, 2, \cdots, n-1). \tag{7c}$$

Again, following Kurth, I shall solve the Newtonian system (7a,c) by successive approximations, but in a modified manner. Note that if $\mu_k = 0$, then (7c) can be solved by concentric circular solutions, with frequencies $\omega_k$ given by **Kepler's Third Law** (1619) as

$$(\omega_k)^2 = G \cdot (1 + \mu_k)/(\rho_k)^3, \qquad \rho_k := \|x^k\|. \tag{8}$$

For the purpose of successive approximations, replace (7c) by the equivalent *ordinary differential equation* (**ODE**)

$$\dot{v}^k = -(\omega_k)^2 x^k + G \cdot \mu_k \cdot \left[ [(x^m - x^k)/\|x^m - x^k\|^3] - [x^m / \|x^m\|^3] \right] + [(\omega_k)^2 - G \cdot (1 + \mu_k)/\|x^k\|^3] \cdot x^k, \tag{9}$$

which makes clear the fact that the system is just a perturbation of the *harmonic oscillator* problem.

We shall need the following lemma, which is an obvious modification of well known basic results in the theory of ODEs [25], [26], [27], [28].

### Lemmas from ODE Theory

<u>DEFINITION</u>. *An m-vector function $f: \mathbb{E}^m \to \mathbb{E}^m$ has a global Lipschitz constant $\kappa$ if there is a positive real number $\kappa > 0$ such that, for all $z^i \in \mathbb{E}^m$,*

$$\| f(z^2) - f(z^1) \| \leq \kappa \cdot \| z^2 - z^1 \|. \tag{10}$$

<u>THEOREM</u>. *Consider the m-vector ODE Initial Condition Problem* (**ICP**)

$$\dot{z} = f(z), \qquad z(0) = z^o, \qquad (0 \leq t \leq T), \tag{11}$$

*where $f$ satisfies a global Lipschitz condition, and where the maximum time $T$ is arbitrary but finite. Let A denote an arbitrary $m \times m$ constant real matrix, and define the vector function $g$ by $g(z) := f(z) - Az$, which also is globally Lipschitzian, with Lipschitz constant now $[\kappa + \|A\|]$,*

where $\|A\|$ denotes the Euclidean norm of A. Now construct the sequence of functions $\{z^k(t)\}$ by successive solution of the forced linear ODE

$$\dot{z}^k = Az^k + g(z^{k-1}(t)), \quad z(0) = z^o, \quad (k = 2, 3, \cdots), \tag{12}$$

where the initial iterate $z^1(t)$ is any arbitrary continuous function of t on [0, T] such that its initial value $z^1(0) = z^o$; or, equivalently, construct the sequence by

$$z^k(t) = exp(At) \cdot z^o + \int_0^t exp(A[t - \tau]) \cdot g(z^{k-1}(\tau)) \, d\tau, \quad 0 \le t \le T, \tag{13}$$

i.e. by integrating the right-hand-side to obtain the left-hand side, and then inserting the result into the right-hand-side and repeating the operation. The resulting sequence is guaranteed to converge uniformly on [0, T] to the exact and unique solution of (11), no matter what is the arbitrary initial iterate $z^1(t)$, and no matter what is the arbitrary matrix A, and no matter how large is the [fixed] time-interval [0,T].

PROOF. Any standard work on ODE's will show how to use Lagrange's method of *variation of constants* to prove that the ODE ICP

$$\dot{z} = Az + g(z), \quad z(0) = z^o, \quad (0 \le t \le T), \tag{14}$$

is completely equivalent to the Volterra integral equation problem

$$z(t) = exp(At) \cdot z^o + \int_0^t exp(A[t - \tau]) \cdot g(z(\tau)) \, d\tau, \quad 0 \le t \le T, \tag{15}$$

where

$$exp(At) := \sum_{k=0}^{\infty} A^k \cdot t^k / k!. \tag{16}$$

Therefore, if the sequence in (12)-(13) converges uniformly, then the result satisfies (15) and so is the solution of (11). Because every continuous function on a closed and bounded subset of a finite-dimensional Euclidean space has (and assumes) a finite maximum and minimum, there exists (on the closed, bounded subset $[0,T] \subset \mathbb{E}^1$)

$$\phi_o = \underset{t \in [0,T]}{MAX}\{ \| z^1(t) - z^o(t) \| \}. \tag{17}$$

By repeatedly applying the triangle inequality to (16) one finds readily that

$$\|exp(At)\| \le \sum_{k=0}^{\infty} \|A\|^k \cdot t^k / k! \equiv exp(\|A\| \cdot t) \le \gamma := exp(\|A\| \cdot T). \tag{18}$$

Accordingly it is easy to prove by induction that (setting $\kappa_1 = \kappa + \|A\|$)

$$\| z^{k+1}(t) - z^k(t) \| \leq \phi_o \cdot (\gamma \cdot \kappa_1 \cdot t)^k / k!, \tag{19}$$

$$\lim_{N \to +\infty} \left\{ \| z^N(t) - z^o(t) \| \right\} \leq \sum_{k=0}^{\infty} \| z^{k+1}(t) - z^k(t) \| \leq \phi_o \cdot exp(\gamma \cdot \kappa_1 \cdot T), \tag{20}$$

for each $t \in [0,T]$, so that $\{ z^k(t) \}$ is a Cauchy sequence (for each $t$) and its limit $z(t)$ must exist. Consideration of the approximating sums $\Sigma_N$ and $\Sigma_{N+1}$ to the series in (20) shows that the convergence is uniform. □

In the sequel, we shall frequently use the 2 × 2 identity matrix $I_2$ and its 'imaginary' skew-symmetric counterpart $J_2 \equiv -(J_2)'$ defined as

$$I_2 = \begin{bmatrix} 1 , & 0 \\ 0 , & 1 \end{bmatrix}, \quad J_2 = \begin{bmatrix} 0 , & 1 \\ -1 , & 0 \end{bmatrix}, \quad (J_2)^2 = -I_2. \tag{21}$$

For present purposes, the result that

$$exp(J_2 \cdot \phi) \equiv \begin{bmatrix} cos(\phi), & sin(\phi) \\ -sin(\phi), & cos(\phi) \end{bmatrix}, \tag{22}$$

will be all-important. (To prove it, insert (21c) into (16).) A similar result which will be needed is that if the 4 × 4 matrix $A$ is defined by (using *MATLAB* notation) $A = [0, I_2; -\omega^2 I_2, 0]$, then it is easy to prove by induction that

$$A^{2k} = (-1)^k \cdot \omega^{2k} \cdot I_4, \quad A^{2k+1} = (-1)^k \cdot \omega^{2k} \cdot A, \tag{23}$$

whence from (16) it is immediate that [letting $I$ denote $I_2$]

$$exp(A \cdot \phi) = exp\begin{bmatrix} 0 , & I \cdot \phi \\ -\omega^2 \cdot I \cdot \phi, & 0 \end{bmatrix} = \begin{bmatrix} cos(\omega \cdot \phi) \cdot I, & [1/\omega] \cdot sin(\omega \cdot \phi) \cdot I \\ -\omega \cdot sin(\omega \cdot \phi) \cdot I, & cos(\omega \cdot \phi) \cdot I \end{bmatrix}. \tag{24}$$

## Principal Result

THEOREM. *Consider the semi-restricted Copernican-Newtonian* $(n + 1)$-*body problem as formulated in* (7a) *and* (9), *with Initial Conditions* (ICs) *corresponding to Keplerian concentric circular orbits, i.e. in which*

$$x^k(0) \cdot v^k(0) = 0, \tag{25a}$$

*so that 3 parameters suffice to specify the ICs. Let these parameters be defined by the triplet* $(\rho,\theta,\omega)$ *given by polar coordinates in which* $\rho$ *denotes the initial orbital radius,* $\theta$ *the initial phase angle, and* $\omega$ *the initial angular velocity, assumed to conform to Kepler's Law* (8), *which* *reduces the IC to a 2-parameter set $(\rho,\theta)$. Specifically, assume (8) and let the ICs of (7a) & (9) be*

$$x^k(0) = \rho_k(cos(\theta_k^o),sin(\theta_k^o))', \quad v^k(0) = \omega_k \cdot J_2 \cdot x^k(0) \equiv - \omega_k \cdot \rho_k(cos(\theta_k^o),sin(\theta_k^o))', \quad (25_b)$$

*where ' denotes vector-matrix transposition. Suppose further that the ICs are so chosen that the frequencies are resonant as in (1); for later convenience, we may assume that*

$$1 \leq m_i, m_j \leq 5, \quad (26)$$

*although the main result holds when these integers are arbitrarily large, which means that the initial frequencies $(\omega_i,\omega_j)$ are essentially arbitrary. Now, keeping the **resonant** frequencies $(\omega_i,\omega_j)$ found at $\mu = 0$ **fixed**, consider the variation of $(\rho,\theta)$ as a function of $\mu$ in such a way as to preserve the periodicity* [namely, "isoperiodic" continuation]:

$$x^i(t + T) \equiv x^i(t), \quad T = 2 \cdot \pi/\omega, \quad \omega = \omega_i/m_i = \omega_j/m_j, \quad (27)$$

*while ignoring the question of periodicity of $x^j$, $j = i + 1$. A necessary and sufficient condition that for sufficiently small $\mu$ there exist ( $\rho_i(\mu)$, $\theta_i(\mu)$ ) preserving the periodicity (27) is that there exist a phase-shift $\phi = \phi(\mu)$ and a distal multiplier $\beta = \beta(\mu)$ such that*

$$\theta_{i+1}(\mu) = \theta_i(\mu) + \phi, \quad (28)$$

$$\rho_{i+1}(\mu) = \beta \cdot \rho_i(\mu), \quad (29)$$

*where, in the second post-Keplerian approximation $(\phi,\beta)$ are given by*

$$\phi = m \cdot (\pi/2) + \cdots, \quad (m = 1, 2, 3, \cdots), \quad (30_a)$$

$$\beta = \beta_o + \cdots \equiv 1/\sqrt{(3/2)^{2/3} - 1} + \cdots = 1.794980 + \cdots, \quad (30_b)$$

*whence, by Kepler's Law in the form*

$$r \equiv m_i/m_{i+1} = \beta^{3/2} + \cdots = 2.40 + \cdots, \quad (30_c)$$

*the ONLY [!] possibilities for the resonance in the low-order case (26) are*

$$m_i:m_{i+1} = 2:1 \text{ or } 5:2 \text{ or } 3:1. \quad (30_d)$$

REMARK. *Cancel the arbitrary **supplementary** assumption (26) which leads to (30d); then the initial generating orbits can be placed arbitrarily closely to **ANY** pair of concentric circles with arbitrary radii (because an arbitrary irrational $\beta$ in (29), and so its $(3/2)^{th}$ power $r$ as in (30c), can be approximated as closely as desired by a rational number $m_i/m_{i+1}$). In this way it can be seen* that the *unique* distal multiplier β given in (30b) as 1.80 at $\mu = 0$ is a **constant** which is completely independent of the assumed trial value of β at $\mu = 0$ !

PROOF. Let $e^i$ denote the columns of $I_2 = (e^1, e^2)$. Then at $\mu = 0$ the solution of the given ODE ICP is

$$x^k(t) = \rho_k \cdot exp(J_2 \cdot \theta_k) \cdot e^1, \quad \theta_k = \omega_k \cdot t + \theta_k^o = m_k \cdot \sigma + \theta_k^o, \quad \sigma := \omega \cdot t. \tag{31}$$

We want this to be the linear part of the reformulation presented in the Lemma, so choose A as in (24), and then re-express the problem in the equivalent form (15), where the required matrix exponential is given by (24); here now $z = (x', v')' \in \mathbb{E}^4$ and, correspondingly, the $g(z)$ in (15) is given by $g = ( 0', [g^{km}]')'$, where

$$g^{km} = G \cdot \mu_k \cdot \left[ [(x^m - x^k)/\|x^m - x^k\|^3] - [x^m/\|x^m\|^3] \right] + [(\omega_k)^2 - G \cdot (1 + \mu_k)/\|x^k\|^3] \cdot x^k. \tag{32}$$

Thus, define the 'osculating' **generating solution** at $\mu = 0$ by

$$x^{k,o}(t) = cos(\omega_k \cdot t) \cdot x^k(0) + [1/\omega_k] \cdot sin(\omega_k \cdot t) \cdot v^k(0), \tag{33a}$$

$$v^{k,o}(t) = cos(\omega_k \cdot t) \cdot v^k(0) - \omega_k \cdot sin(\omega_k \cdot t) \cdot x^k(0). \tag{33b}$$

Now the problem is rigorously equivalent to solving the integral equation

$$x^k(t) = x^{k,o}(t) + \int_0^t [1/\omega_k] \cdot sin(\omega_k[t - \tau]) \cdot g^{km}(x^k(\tau), x^m(\tau)) \, d\tau; \tag{34a}$$

although the formulation (24) gives a second equation

$$v^k(t) = v^{k,o}(t) + \int_0^t cos(\omega_k[t - \tau]) \cdot g^{km}(x^k(\tau), x^m(\tau)) \, d\tau, \tag{34b}$$

the second is a consequence of differentiating the first with respect to time *t*, and therefore is of no further consequence, after noting that in (33a) the initial velocity $v^k(0) = \omega_k \cdot J_2 \cdot x^k(0)$ is already defined in terms of $x^k(0)$ by hypothesis.

It is a well-known consequence of basic ODE theory that the (necessarily unique) solutions of (7a) and (9) for $\mu = 0$, namely (33), and the solutions for sufficiently small $\mu > 0$ remain arbitrarily close together for any finite time, specifically here for $0 \leq t \leq T = 2 \cdot \pi/\omega$, provided only that μ be sufficiently small. Therefore for sufficiently small μ one knows *a priori* that the solutions remain within planar concentric annuli surrounding the circular orbits (33), and therefore do not approach each other during the time of interest. Accordingly one may find a global Lipschitz constant for such relevant domains of the Cartesian products of $\mathbb{E}^4$ (in which the Jacobian matrix of the right-hand side of (11) is continuous, and so has a bounded norm $\kappa$ in the relevant $z$-domain). Therefore the Lemma is applicable, for sufficiently small $\mu$, and so the solution of (34) exists and can be constructed by successive approximations (with $j = 1, 2, 3, \cdots$) :

$$x^{k,j}(t) = x^{k,o}(t) + \int_0^t [1/\omega_k] \cdot sin(\omega_k[t - \tau]) \cdot g^{km}(x^{k,j-1}(\tau), x^{m,j-1}(\tau)) \ d\tau, \qquad (35)$$

taking as the initial iterate the decoupled Keplerian solution (33), for $0 \leq t \leq T$.

Next, look at the solution of (34) only at every revolution of duration $T = 2\cdot\pi/\omega$, wherein the *commensurability* of $\omega_i$ and $\omega_{i+1}$ ensures that there is a **common period** to the initial iterates of the two adjacent orbits. This give the famous **Poincaré map**:

$$x^k(T) = x^k(0) - \mu \cdot (G\mu_m^o/[m_k\omega^2]) \cdot \int_0^{2\pi} sin(m_k\sigma) \cdot h^{km}(x^k(\sigma), x^m(\sigma)) \ d\sigma, \qquad (36a)$$

$$h^{km} := \left[ [(x^m - x^k)/\|x^m - x^k\|^3] - [x^m/\|x^m\|^3] \right] + [(\omega_k)^2 - Gp\cdot(1 + \mu_k)/\|x^k\|^3] \cdot x^k. \qquad (36b)$$

We shall prove that this map can be expressed in the form

$$x^k(T) = x^k(0) - \mu \cdot (G\mu_m^o/[m_k\omega^2]) \cdot f^{km}(\rho_k, \theta_k^o, \rho_m, \theta_m^o; \mu), \qquad (37a)$$

$$f^{km}(\rho_k, \theta_k^o, \rho_m, \theta_m^o; \mu) := \int_0^{2\pi} sin(m_k\sigma) \cdot h^{km}(x^k(\sigma), x^m(\sigma)) \ d\sigma, \qquad (37b)$$

where for certain specific values of $(\rho_k, \theta_k^o)$ the Jacobian determinant of $f^{km}$ with respect to $(\rho_k, \theta_k^o)$ is positive at $\mu = 0$, and where $f^{km} = 0$ for these same specific values, at $\mu = 0$; then, by the Implicit Function Theorem there exist functions $(\rho_k(\mu), \theta_k(\mu))$ for sufficiently small values of $\mu$ which satisfy $f^{km} \equiv 0$ and so which provide the desired periodicity of $x^k(T)$.

It will become evident that the conditions to be derived are also necessary, because if they are not satisfied, $f^{km}$ provides a *resonant* forcing term which drives $x^k(T)$ ever farther from $x^k(0)$, and will do so without bound (or until the resonant 'pumping' drives the map's next iterate out of the map's domain of definition).

To anticipate the results of a rather lengthy and arduous calculation, it will be proved below that $f^{km}$ has the form

$$f^{km} = \eta_{km,m} \cdot exp(J_2\Phi_{km,m}) \cdot x^m(0) - \eta_{km,k} \cdot exp(J_2\Phi_{km,k}) \cdot x^k(0), \qquad (38)$$

where the scalar functions $\eta_{kmj}(\mu)$ and $\Phi_{kmj}(\mu)$ have the properties that

$$\eta_{km,m}(0) \;=\; 0, \quad \eta_{km,k}(0) \;=\; 1/2, \quad \Phi_{km,k}(0) \;=\; \pi/2, \tag{39}$$

so that at $\mu = 0$ the Jacobian matrix $H$ of $f^{km}$ with respect to $x^k(0)$ is given by

$$H \;=\; -(1/2)\cdot exp(J_2\cdot[\pi/2]) \;=\; -(1/2)\cdot J_2, \quad det(H) \;=\; 1/4 \;>\; 0, \tag{40}$$

and, as claimed, the **Leray-Schauder Index** [29] of the chosen generating solution is unity.

It is evident by inspection that upon the second iteration the third term (36b) drops out by the choice of a Keplerian generating solution, so that for consideration of the second iterate of the successive approximations we may simplify (36b) to

$$h^{km} \;:=\; [(x^m - x^k)/\|x^m - x^k\|^3] \;-\; [x^m/\|x^m\|^3]. \tag{41}$$

It is also immediate that the second term in (41) may be omitted, because by (31)

$$x^m/\|x^m\|^3 \;=\; (1/\rho_m^2)\cdot exp(J_2\cdot m_m\sigma)\cdot exp(J_2\cdot\theta_m^o)\cdot e^1, \quad \int_0^{2\pi} sin(m_k\sigma)\cdot exp(J_2\cdot m_m\sigma)\,d\sigma \;\equiv\; 0 \tag{42}$$

because of the hypothesis that $m_k \neq m_m$. Thus we have simplified the calculation to that of evaluation of

$$f^{km} \;=\; \int_0^{2\pi} sin(m_k\sigma)\cdot\Psi(\sigma)\cdot(x^m(\sigma) - x^k(\sigma))\,d\sigma, \quad x^k(\sigma) \;=\; \rho_k\cdot exp(J_2\cdot[m_k\sigma + \theta_k^o])\cdot e^1, \tag{43}$$

$$\Psi(\sigma) \;:=\; 1/\|x^m(\sigma) - x^k(\sigma)\|^3. \tag{44}$$

Now by the chief property of the scalar product $(x,y)$, namely that if $M$ is an arbitrary matrix $(Mx,y) \equiv (x,M'y)$, and the fact that $J_2' = -J_2$, it is easy to calculate that

$$\| x^m(\sigma) - x^k(\sigma) \|^2 \;\equiv\; \|x^m\|^2 + \|x^k\|^2 - 2\cdot\rho_m\cdot\rho_k(e^1, exp(J_2\xi)e^1), \tag{45a}$$

$$\xi \;=\; (m_k - m_m)\cdot\sigma + \theta_k^o - \theta_m^o, \tag{45b}$$

$$\Psi(\sigma) \;=\; \psi(\sigma)/[\,(\rho_k)^2 + (\rho_m)^2\,]^{3/2}, \tag{45c}$$

$$\psi(\sigma) \;=\; 1/[\,1 - (\varepsilon_{km})^2 cos(\xi)\,]^{3/2}, \quad (\varepsilon_{km})^2 \;:=\; 2\cdot\rho_k\cdot\rho_m/[\,(\rho_k)^2 + (\rho_m)^2\,], \tag{45d}$$

$$\psi(\sigma) \;=\; 1 + (3/2)\cdot(\varepsilon_{km})^2 cos(\xi) + (15/8)\cdot(\varepsilon_{km})^4 cos^2(\xi) + \cdots, \tag{45e}$$

where the binomial series in (45e) always converges because $(\varepsilon_{km})^2 < 1$ whenever $\rho_k$ and $\rho_m$ are distinct (which is a consequence of the fact that then $0 < (\rho_k - \rho_m)^2$ and of obvious manipulations of the expansion of the latter).

- 11 -

Now, remembering that $\psi$ depends upon both $k$ and $m$, define

$$A_{kmj} := (1/[2\pi]) \cdot \int_0^{2\pi} sin(m_k \sigma) \cdot cos(m_j \sigma) \cdot \psi(\sigma) \, d\sigma, \tag{46a}$$

$$B_{kmj} := (1/[2\pi]) \cdot \int_0^{2\pi} sin(m_k \sigma) \cdot sin(m_j \sigma) \cdot \psi(\sigma) \, d\sigma, \tag{46b}$$

$$\eta_{kmj} := \{ (A_{km})^2 + (B_{km})^2 \}^{1/2}, \tag{46c}$$

$$\Phi_{kmj} := -Arctan\{ B_{kmj}/A_{kmj} \}, \tag{46d}$$

and note that (43)-(44) may be simplified by use of the novel identity

$$(1/[2\pi]) \cdot \int_0^{2\pi} sin(m_k \sigma) \cdot \psi(\sigma) \cdot exp(J_2[m_j \sigma + \theta_j^o]) \, d\sigma \equiv \eta_{kmj} exp(J_2[\Phi_{mkj} + \theta_j^o]). \tag{47}$$

It is the radical simplification provided by the apparently hitherto unnoticed identity (47) which appears to be the chief innovation in the present work. For we may now write that

$$f^{km} = d^{km}/[ (\rho_k)^2 + (\rho_m)^2 ]^{3/2}, \tag{48a}$$

$$d^{km} := \left[ \rho_m \eta_{kmm} \cdot exp(J_2[\Phi_{kmm} + \theta_m^o]) - \rho_k \eta_{kmk} \cdot exp(J_2[\Phi_{kmk} + \theta_k^o]) \right] \cdot e^1 \equiv$$

$$\equiv \rho_m \cdot \eta_{kmm} \cdot \begin{pmatrix} cos(\Phi_{kmm} + \theta_m^o) \\ -sin(\Phi_{kmm} + \theta_m^o) \end{pmatrix} - \rho_k \cdot \eta_{kmk} \cdot \begin{pmatrix} cos(\Phi_{kmk} + \theta_k^o) \\ -sin(\Phi_{kmk} + \theta_k^o) \end{pmatrix}. \tag{48b}$$

From mere inspection of (48) it is now evident that the *necessary and sufficient conditions* for $f^{km}$ to vanish are that there be an **orbital resonance** defined by

$$\rho_m = \beta \cdot \rho_k, \quad \beta = \eta_{kmk}/\eta_{kmm}, \tag{49a}$$

and a corresponding enabling phase-shift

$$\theta_m^o = \theta_k^o + \phi_{mk} + M \cdot \pi, \quad \phi_{mk} = \Phi_{kmk} - \Phi_{kmm}, \quad ( M = 1, 2, 3, \cdots ). \tag{49b}$$

This completes the easy part of the present derivation.

Now begins the hard work of evaluation of $\beta$ and $\phi$. Part of this is easy, because by inspection we need only to evaluate the lowest-order terms in

$$A_{kmk} = 0 + \cdots, \quad B_{kmk} = (1/2) + \cdots, \tag{50}$$

because in this case it is adequate to use $\psi = 1 + \cdots$ because of the well known orthogonality properties of sines and cosines. In contrast, the lowest order terms in $A_{kmm}$ and $B_{kmm}$ vanish identically, and we must go to the second-order terms in $\psi$ in order to get meaningful results. Thus, to the second order in the series (45e)

$$(A_{kmm}, B_{kmm}) = (3/2) \cdot (\varepsilon_{km})^2 \cdot (\alpha_{kmm}, \beta_{kmm}) + \cdots, \tag{51a}$$

$$\alpha_{kmm} := (1/[2\pi]) \cdot \int_0^{2\pi} \sin(m_k \sigma) \cdot \cos(m_m \sigma) \cdot \cos([m_k - m_m] \sigma + \theta_k^o - \theta_m^o) \, d\sigma, \tag{51b}$$

$$\beta_{kmm} := (1/[2\pi]) \cdot \int_0^{2\pi} \sin(m_k \sigma) \cdot \sin(m_m \sigma) \cdot \cos([m_k - m_m] \sigma + \theta_k^o - \theta_m^o) \, d\sigma, \tag{51c}$$

which is where conceptualizing ends and labor begins. My advice to the reader is to replace each sine and cosine by the sum or difference of two complex exponentials, as in de Moivre's Theorem, and then multiply out the resulting 6 products as complex numbers. From this lengthy exercise in elementary complex algebra there results the [ultimately real] *identities:*

$$\alpha_{kmm} = -(1/4) \cdot \sin(\theta_k^o - \theta_m^o), \qquad \beta_{kmm} = (1/4) \cdot \cos(\theta_k^o - \theta_m^o), \tag{51d}$$

from which we obtain the welcome simplification that $[(\alpha_{kmm})^2 + (\beta_{kmm})^2]^{1/2} \equiv 1/4$. Therefore, finally, by (46c)

$$\eta_{kmm} = (3/2) \cdot (\varepsilon_{km})^2 \cdot (1/4) + \cdots, \qquad \eta_{kmk} = (1/2) + \cdots, \tag{51e}$$

so that, by (49a), to lowest order in $\mu$,

$$1/\beta = \rho_k/\rho_m = \eta_{kmm}/\eta_{kmk} = (3/2) \cdot \rho_k \cdot \rho_m / [(\rho_k)^2 + (\rho_m)^2]^{3/2} \equiv$$

$$\equiv (3/2) \cdot (1/\beta) / [1 + (1/\beta)^2]^{3/2}, \tag{51f}$$

which *requires* for self-consistency that, reminiscent of the **1766/1772 Titius/Bode 'Law'**,

$$1 + (1/\beta)^2 = (3/2)^{2/3}, \tag{51g}$$

i.e. that $\beta$ have the *unique* particular value claimed in (30b)! Note that to lowest order this $\beta$ is a *universal constant*, **independent** of the gravitational constant $G$ or the masses of the planets!

The proof of (30a) is analogous but simpler, noting that from (46d) and (51d)

$$\phi_{mk} = \Phi_{kmk} - \Phi_{kmm} = -(\theta_m^o - \theta_k^o), \tag{51h}$$

so that, bringing $(\theta_m^o - \theta_k^o)$ to the left-hand side of (49b) and dividing by 2 we obtain the claimed result (30a). Equation (38) is a trivial consequence of (48b). Also evaluation of the Jacobian with respect to $(\rho, \theta)$ instead of with respect to the components of makes $x(0)$ makes no difference to the claim that the Leray-Schauder Index of the generating solution isolated by (51g) is unity. This completes the proof.

Now that the importance of the distal multiplier $\beta$ for the distance beyond an inner planet for an outer planet in terms of the ratio of orbital radii has been derived in full rigor for the coplanar, concentric 3-body problem, it seems permissible to use a cruder physical model to consider the case of 4 or more bodies. For simplicity, keep the first body anchored at the origin and let the bodies have mean orbital radii $\rho_i$ ($i = 1, 2, \cdots n$), where $\rho_1 \equiv 0$ by definition, and

$$0 = \rho_1 < \rho_2 < \cdots < \rho_i < \rho_{i+1} < \cdots < \rho_n, \tag{55}$$

and where each body (initially assumed decoupled from mutual interactions) is started with Keplerian circular velocity $v_i = \sqrt{G \cdot M}/\sqrt{\rho_i}$ where $M = m_1$ denotes the large central mass and where the smaller masses are denoted by $m_i = \mu_i \cdot M$ in terms of ratios $\mu_i \ll 1$ ($i = 2, 3, \cdots, n$). Let $\Gamma_0 > 0$ denote initial total angular momentum, and $E_0 = |\mathcal{E}_0| = -\mathcal{E}_0 > 0$ denote the absolute value of the initial total energy. Also, define $\gamma_0 \equiv \Gamma_0/\sqrt{G} \cdot M^{3/2}$ and $\varepsilon_0 \equiv 2 \cdot E_0/\sqrt{G} \cdot M^2$, and it is easy to verify that (neglecting terms quadratic or higher in the $\mu_i$) Conservation of Angular Momentum and Conservation of Energy (combined with the Virial Theorem) are given by $\gamma = \gamma_0$ and $\varepsilon = \varepsilon_0$ where

$$\gamma = \sum_{i=2}^{n} \mu_i \sqrt{\rho_i}, \qquad \varepsilon = \sum_{i=2}^{n} \mu_i/\rho_i, \tag{56}$$

<u>THEOREM</u>.  *If the planetary orbits satisfy a Titius-Bode Law of the form*

$$\rho_i = \beta^{i-2} \cdot \rho_2, \qquad (i = 2, 3, 4, \cdots, n), \tag{57}$$

*then the distal ratio $\beta = z^2$ is the square of the unique positive root $z > 0$ of a polynomial of degree $4 \cdot n - 8$ in $z$ whose coefficients are functions only of the constants $\gamma_0$ and $\varepsilon_0$ and the mass-ratios $\mu_i$ and which can be derived by elimination of $\rho_2$ between the expressions for $\gamma$ and $\varepsilon$ in (56).*

<u>PROOF</u>. Insert (57) into (56) and compute $\gamma^2 \cdot \varepsilon$. I shall publish a general algorithm defining all of the coefficients explicitly elsewhere; however, it can be recovered easily by the interested reader after following the next example in the case $n = 3$, wherein the algebra is less difficult. □

<u>REMARK</u>. To apply this result to the present solar system, simply replace the $\rho_i$ in (56) by the actual mean distances $R_i$ from observational astronomy and use the actual mass-ratios $\mu_i$ from astrophysics in order to evaluate $\gamma_0$ and $\varepsilon_0$ numerically. After finding $\beta$, the eliminated $\rho_2$ can be recovered by solving either $\gamma = \gamma_0$ or $\varepsilon = \varepsilon_0$, following which the remaining $\rho_i$ are given by (57).

In the case $n = 3$, the polynomial $F(z)$ whose roots are sought has the form

$$F(z) = z^4 + \alpha_3 \cdot z^3 + \alpha_2 \cdot z^2 + \alpha_1 \cdot z^3 + \alpha_0, \tag{58}$$

which, as in the proof of the preceding theorem, can be obtained simply by comparing coefficients in the multiplied-out version of

$$(\mu_1 \cdot z^2 + \mu_2) \cdot (\mu_1 + \mu_2 \cdot z)^2 - \alpha \cdot z^2 = 0, \qquad \alpha = \varepsilon_0 \cdot (\gamma_0)^2. \tag{59}$$

The result is

$$\alpha_3 = 2 \cdot (\mu_1/\mu_2), \quad \alpha_2 = (\mu_1/\mu_2)^2 + (\mu_2/\mu_1) - \alpha/[\mu_1(\mu_2)^2], \quad \alpha_1 = 2, \quad \alpha_0 = \mu_1/\mu_2. \tag{60}$$

Using data for Jupiter and Saturn, the reader may verify that $F = 0$ has only one positive root, whosee square yields $\beta = 1.833$. With more effort, analogous results can be obtained for arbitrary *n (see Table 1 ). Note that the value of b had converged to 1.795, in amazing agreement with the value obtained in (30b).*

| $m_i$ | $R_i$ | $\rho_i$ | **BODY** | $n$ | $\beta$ | $\beta_i$ |
|---|---|---|---|---|---|---|
| 1,000,000 | | | Sun | 1 | | |
| 0.16601 | 0.3871 | 0.2818 | +Mercury | 2 | | |
| 2.447841 | 0.7233 | 0.50608 | +Venus | 3 | 1.86 | 1.868 |
| 3.003469 | 1.000 | 0.9088 | +Earth | 4 | 1.46 | 1.382 |
| 0.322714 | 1.5237 | 1.632 | +Mars | 5 | 1.47 | 1.5237 |
| A | | | | 6 | | |
| 954.60258 | 5.2030 | 5.26 | +Jupiter | 7 | 1.68 | 1.927 |
| 285.80765 | 9.5281 | 9.45 | +Saturn | 8 | 1.766 | 1.831 |
| 43.549846 | 19.1829 | 16.97 | +Uranus | 9 | 1.800 | 2.013 |
| 51.67161 | 30.0796 | 30.48 | +Neptune | 10 | 1.795 | 1.568 |
| 0.007541 | 49.0250 | 54.70 | +Pluto | 11 | 1.795 | 1.6298 |

**Table 1**

## Conclusions

In this paper, we showed that in a planetary or satellite system, resonant perturbations will cause dynamical evolution of the orbital radii except under highly specific mutual relationships. Simplifying the problem by reformulating it as a hierarchical sequence of [unrestricted] 3-body problems, in which gravitational interactions are ignored except between the central body and the body of interest and the next outwardly orbitally adjacent body, it is proved that the resonant perturbations from the outer body will destabilize the inner body [& conversely] unless its mean orbital radius is a unique and specific multiple $\beta$ of that of the inner body . In this way a sequence of concentric orbits can each stabilize the next adacent inner orbit, until only the outermost orbit remains; but it is already tied to the collection of inner orbits by conservation of total angular momentum and so the entire configuration becomes stabilized. Let $\mu = M_{\text{Max}}/M$ denote the ratio of the mass of the largest small body $M_{\text{MAX}}$ to that, M,of the large central body; in our Solar System, $\mu$ is less than $10^{-3}$. Expanding $\beta$ in a power series in $\mu$, the lowest-order terms for distal multiplier $\beta$ and phase shift $\phi$ are found to start with the following universal constants [for $m = 1, 2, 3$]

$$\beta \equiv \beta + ...... = 1/\sqrt{(3/2)^{(2/3)} - 1} = 1.794980 + ....., \phi = m(\pi/2) + ..$$

which agrees with the observed distal ratios between Jupiter and the Asteroids, and between Saturn & Jupiter, and Uranus & Saturn.

## Acknowledgements

For primary encouragement, which led me to my first foaray [39] in this subject, I am deeply indebted to Dr. C.D. Johnson and his former student S. Addington. Also I am indebted to Dr. Miklos Farkas [15], to Dr. J. E. Chambers [32], and to Dr. P. Morrison [31], for their interest and for valuable literature references.